\documentstyle[12pt]{article}

\begin{document}


\title{Spinless Matter in Transposed-Equi-Affine Theory of Gravity
\thanks{Directed to the journal GENERAL RELATIVITY AND GRAVITATION}}

\author{
{\normalsize P.~P.~Fiziev}\\
{\footnotesize Department of Theoretical Physics, Faculty of Physics,}\\
{\footnotesize Sofia University,}\\
{\footnotesize Boulevard~5 James Bourchier, Sofia~1164, }\\
{\footnotesize Bulgaria }\\
\\
{\footnotesize  E-mail: fiziev@phys.uni-sofia.bg}
}

\maketitle

\begin{abstract}
We derive and discus the equations of motion for spinless matter:
relativistic spinless scalar fields, particles and fluids in the recently
proposed by A. Saa model of gravity with covariantly constant volume
with respect to the transposed connection in Einstein-Cartan spaces.

A new interpretation of this theory
as a theory with variable Plank "constant"
is suggested.

We show that the consistency of the semiclassical limit of the wave
equation and classical motion dictates a new definite universal interaction
of torsion with massive fields.

\end{abstract}


\sloppy
\renewcommand{\baselinestretch}{1.3} %
\newcommand{\sla}[1]{{\hspace{1pt}/\!\!\!\hspace{-.5pt}#1\,\,\,}\!\!}
\newcommand{\db}{\,\,{\bar {}\!\!d}\!\,\hspace{0.5pt}}
\newcommand{\lambdab}{\,\,{\bar {}\!\!\lambda}\!\,\hspace{0.5pt}}
\newcommand{\partb}{\,\,{\bar {}\!\!\!\partial}\!\,\hspace{0.5pt}}
\newcommand{\dsla}{\partb}
\newcommand{\eql}{e _{q \leftarrow x}}
\newcommand{\eqr}{e _{q \rightarrow x}}
\newcommand{\ite}{\int^{t}_{t_1}}
\newcommand{\itz}{\int^{t_2}_{t_1}}
\newcommand{\itd}{\int^{t_2}_{t}}
\newcommand{\lfrac}[2]{{#1}/{#2}}
\newcommand{\sfrac}[2]{{\small \,\,\hbox{${\frac {#1} {#2}}$}}}
\newcommand{\dV}{d^4V\!\!ol}
\newcommand{\ben}{\begin{eqnarray}}
\newcommand{\een}{\end{eqnarray}}
\newcommand{\la}{\label}


\section{Introduction}
The Einstein-Cartan theory of gravity (ECTG) has a long history  --
see for example the review articles \cite{Hehl1}--\cite{Hehl3} and the huge
amount of references therein.
Despite of the obvious beauty of this theory and of the fundamental
physical and geometrical ideas on which it was built
there exist some long standing and well known problems in it.

In present article we consider one of them --
the discrepancy between the results obtained via the use of the
minimal coupling principle (MCP) in the action principle and directly
in the equations of motion:
if one substitutes the covariant derivatives
into the special relativistic equations of motion for flat space
one reaches a result which differs from the one
obtained when one substitutes the covariant derivatives
in the action functional and then derives the equations of motion
from {\em standard} action principle.

In the standard version of ECTG the usual variational
principle is used after applying MCP in the action integral \cite{Hehl1}.
Then the interaction of the fields with different spins
with torsion does not obey the MCP at the level of the equations of motion
\cite{Hehl1}, \cite{Hehl4}.
This is equivalent to introduction of a strange
torsion-force-like-terms in the equations of motion
in different way for different spins.
As a result the equivalence principle is violated.

To be specific, let us consider the simplest case of the spinless matter.

For scalar field $\phi(x)$ with mass $m$ in the standard ECTG the MCP produces
the action
\ben
{\cal A}[\phi(x)]= \int d^4x \sqrt{|g(x)|}  {\sfrac 1 2}\!
\left( g^{\mu \nu}(x) \nabla_\mu \phi(x) \nabla_\nu \phi(x)
- m^2 \phi^2(x) \right)
\la{sfA}
\een
in the four-dimensional Einstein-Cartan space
${\cal M}^{(1,3)}\{g_{\alpha\beta}(x), \Gamma^\gamma_{\alpha\beta}(x)\}$,\,\,
$g_{\alpha\beta}(x)$ being the metric tensor with signature (+,-,-,-),\,\,
$\Gamma^\gamma_{\alpha\beta}(x)$ being the coefficients
of the metric compatible connection: $\nabla_\alpha g_{\beta \gamma}\equiv 0$,
$\nabla_\alpha$ being the covariant derivative with respect to the affine
connection with coefficients
$\Gamma^\gamma_{\alpha\beta}={\gamma \brace \alpha \beta}
- {K_{\alpha\beta} }^\gamma$, torsion
${S_{\alpha\beta} }^\gamma =  \Gamma^\gamma_{[\alpha\beta]}$,
and contorsion ${K_{\alpha\beta} }^\gamma = - {S_{\alpha\beta} }^\gamma
- {S^\gamma}_{\alpha\beta} - {S^\gamma}_{\beta\alpha}$;
${\gamma \brace \alpha \beta}= {\sfrac 1 2}g^{\gamma \mu}
(\partial_\alpha g_{\mu\beta} +\partial_\beta g_{\mu\alpha} -
\partial_\mu g_{\alpha\beta})$ being the Christoffel symbols.

For scalar field $\nabla_\alpha \phi \equiv \partial_\alpha \phi $
and performing standard variation of the action (\ref{sfA}) we reach
the equation of motion:
\ben
{\stackrel{{}^{\{\}}}{\Box}} \phi + m^2 \phi =
\Box \phi + m^2 \phi + 3 S^\mu \nabla_\mu \phi = 0.
\la{sfEG}
\een
Here the trace of the torsion $S_\alpha = {\sfrac 2 3}{S_{\alpha \mu}}^\mu$
gives the torsion vector according to the notations of the reference
\cite{Schouten} which we shall use further.
In addition we use the relation
$\Box \phi=
{\stackrel{{}^{\{\}}}{\Box}} \phi - 3 S^\mu \partial_\mu \phi$
between the laplasian $\Box = g^{\alpha\beta} \nabla_\alpha \nabla_\beta$
and Laplas-Beltrami operator
${\stackrel{{}^{\{\}}}{\Box}}=
g^{\alpha\beta} \stackrel{{\{\}}}{\nabla}_\alpha \stackrel{{\{\}}}{\nabla}_\beta
={1\over \sqrt{|g|}}\partial_\mu\left(\sqrt{|g|}g^{\mu\nu}\partial_\nu\right)$
in the space
${\cal M}^{(1,3)}\{g_{\alpha\beta}(x), \Gamma^\gamma_{\alpha\beta}(x)\}$,
$\stackrel{{\{\}}}{\nabla}_\alpha$ being the covariant derivative with
respect to the Levi-Cevita connection with coefficients
${\gamma \brace \alpha \beta}$\footnote{We shall use the mark $\{\}$
above the symbols to denote all objects: operators, quantities, etc.
which correspond to the Levi-Cevita connection in the space
${\cal M}^{(1,3)}\{g_{\alpha\beta}(x), \Gamma^\gamma_{\alpha\beta}(x)\}$.}.

If we consider the affine connection as a fundamental object which
defines the very geometry of the space
${\cal M}^{(1,3)}\{g_{\alpha\beta}(x), \Gamma^\gamma_{\alpha\beta}(x)\}$,
all equations of motion have to be written in terms of its absolute
derivatives. Then the third term $3 S^\mu \nabla_\mu \phi$
in the corresponding form of the equation (\ref{sfEG}) has to be considered
as an additional force-like term, due to the torsion. It has to be introduced
to compensate the natural torsion dependence of the scalar field dynamics
generated by the direct application of the MCP
to the special relativistic equation of motion of spinless field.
The last procedure would lead to the equation of motion
of scalar field in the space
${\cal M}^{(1,3)}\{g_{\alpha\beta}(x), \Gamma^\gamma_{\alpha\beta}(x)\}$
which reads:
\ben
\Box \phi + m^2 \phi = 0.
\la{sfEA}
\een

One has to confess that the treatment of the equation (\ref{sfEG})
in the framework of the basic affine geometry of the space
${\cal M}^{(1,3)}\{g_{\alpha\beta}(x), \Gamma^\gamma_{\alpha\beta}(x)\}$
is quite unnatural. Of course, one may argue, that the pure metric geometry
has got equal rights in the Einstein-Cartan spaces. But it seems to us
that the use of the Levi-Cevita connection in the equations of motion
in ECTG will be a step away from the basic philosophy of this theory.

The same happens for test spinless classical point particles in standard ECTG.
According to MCP the action for test spinless particle with mass $m$
in usual notations acquires the form
\ben
{\cal A}[x(t)]=
-mc\int \sqrt{g_{\mu \nu}(x(t)) \dot x^\mu(t) \dot x^\nu(t)}\,dt=
-mc\int ds.
\la{spA}
\een
Now, the standard action principle leads to the {\em geodesic}
equations of motion
\ben
mc^2\left( {\frac {d^2 x^\gamma} {ds^2}} + { \gamma \brace \alpha \beta}
{\frac {d x^\alpha} {ds}}{\frac {d x^\beta} {ds}} \right) =
mc^2 {\frac D {ds}}{\frac {d x^\gamma} {ds}}
- 2 mc^2 {S^\gamma}_{\alpha \beta}
{\frac {d x^\alpha} {ds}}{\frac {d x^\beta} {ds}} = 0.
\la{spEG}
\een
But the direct application of the MCP to the special relativistic equations
of motion of a test particle leads to {\em autoparallel} equations
in the space
${\cal M}^{(1,3)}\{g_{\alpha\beta}(x), \Gamma^\gamma_{\alpha\beta}(x)\}$:
\ben
mc^2\left( {\frac {d^2 x^\gamma} {ds^2}} +  \Gamma_{\alpha \beta}^\gamma
{\frac {d x^\alpha} {ds}}{\frac {d x^\beta} {ds}} \right)=
mc^2 {\frac D {ds}}{\frac {d x^\gamma} {ds}}  = 0,
\la{spEA}
\een
where ${\frac D {ds}}$ is the absolute derivative with respect to the
affine connection.

Obviously the autoparallel equation (\ref{spEA}) means
a free motion of the test spinless  particle in the space
${\cal M}^{(1,3)}\{g_{\alpha\beta}(x), \Gamma^\gamma_{\alpha\beta}(x)\}$
with zero absolute acceleration:
$a^\gamma = c^2 {\frac D {ds}}{\frac {d x^\gamma} {ds}} = 0$.
This is the most natural translation of the usual dynamics of a
test free particle to the ECTG
and corresponds to the very physical notion of a "free test particle".

In contrast, the geodesic equations (\ref{spEG}) imply the unnatural law
of free motion:\,\, $m a^\gamma = {\cal F^\gamma}$.
Hence, we actually introduce a specific "torsion force"
${\cal F^\gamma} = 2 mc^2 {S^\gamma}_{\alpha \beta} u^\alpha u^\beta$
($u^\alpha = {\frac {d x^\alpha} {ds}}$ being the particle's four-velocity)
to compensate the natural torsion dependence of the dynamics in the space
${\cal M}^{(1,3)}\{g_{\alpha\beta}(x), \Gamma^\gamma_{\alpha\beta}(x)\}$
and to allow the free test particle to follow
the usual extreme of the classical action (\ref{spA}).

We shall call the relations like (\ref{sfEG}) and (\ref{spEG})
{\em equations of geodesic type}, and
the relations like (\ref{sfEA}) and (\ref{spEA}) --
{\em equations of autoparallel type}.

The above paradox in description of the free motion of test particles
forces one to make a choice what to consider as a more fundamental:

1)the free motion as a motion without external forces of any nature, and hence,
with zero absolute acceleration; or

2)the free motion as a motion on geodesic lines
with extremely length, according to the {\em standard} action principle.

It is quite obvious that the first alternative has a more profound physical
character. The only argument to chose the second one is the fact, that the
action principle follows from quantum mechanics as a fundamental principle
for classical motion \cite{Dirac},\cite{Feynman}.
But there is no guarantee that the quantum
mechanics leads to the usual form of action principle in affine connected
spaces with nonzero torsion.
Moreover it is found that Feynman path integral leads to the
Schr\"odinger equation of autoparallel type in such spaces \cite{Kl1}, \cite{F2}.
We shall not give here the derivation of the right variational
principle in general affine connected spaces from quantum mechanics.
Instead in the present article we accept and investigate
the first alternative following other reasons.

The autoparallel motion of test particle in ECTG was
proposed in \cite{Ponamarev} and derived from formally
modified variational principle as early as in \cite{Timan}\footnote{We
ware not able to find the original Timan's article and refer to it following
the second article in the reference \cite{Timan}.}.
One has to add that in Weitzenb\"ock affine flat spaces with torsion
a new variational principle for classical particle trajectories was found
recently \cite{F3},\cite{F4},\cite{F1}.
It leads after all to autoparallel motion of the particles and gives
a proper development of the concept of "quantum equivalence principle"
\cite{Kl1}, \cite{Kl2}, \cite{Kl6}.
Once more a formal modification of the variational principle in spaces with
torsion was examined in \cite{Kl4}.
Very recently the autoparallel motion of nonrelativistic particle was derived
from proper generalization of the Gauss' principle of least constraint
in \cite{Kl3}.

Nevertheless, at present one can't exclude the possibility for geodesic motion.
Therefore we shall take into account this type of motion
for spinless matter, too. The point is to develop both conceptual possibilities
to the form which will admit a comparison with the experimental evidences,
or will recover their theoretical (in)consistency.

For example, there exist the following consistency problem in the affine
connected space ${\cal M}^{(1,3)}\{g_{\alpha\beta}(x),
\Gamma^\gamma_{\alpha\beta}(x)\}$. In the Riemannian space the geodesic equation (\ref{spEG}) for test particles
with mass $m$ follows from the scalar field
equation (\ref{sfEG}) with the same mass in a semiclassical limit,
See for example \cite{Schr}.
One expects to see the same property in the case of nonzero
torsion in the space
${\cal M}^{(1,3)}\{g_{\alpha\beta}(x), \Gamma^\gamma_{\alpha\beta}(x)\}$, too.
But the naive generalization of the corresponding procedure does not lead to
the expected result.
Indeed, representing the field $\phi$ in a form $\phi= A \exp(i\varphi)$
with some real amplitude $A$ and real phase $\varphi$ we can write down
$\Box \phi = \phi\left( {\Box A \over A} -
g^{\alpha\beta} \partial_\alpha \varphi \partial_\beta \varphi \right)+
i{\phi\over A^2}
\nabla_\alpha \left(A^2 g^{\alpha\beta} \nabla_\beta \varphi \right)$.
Now the autoparallel equation (\ref{sfEA}) in the semiclassical limit
${\Box A \over A}\approx 0$ yields the eikonal equation
\ben
g^{\alpha\beta} \partial_\alpha \varphi \partial_\beta \varphi = m^2
\la{EIC}
\een
which seems to correspond to the Hamilton-Jacobi equation for classical
action function $S=\hbar \varphi$ of the {\em geodesic} equation (\ref{spEG}),
not of the autoparallel one (\ref{spEA}),
See for details Section 7 of the present article.
In addition we reach the autoparallel type of conservation law
$\nabla_\alpha j^\alpha=0$
for the current $j_\alpha = A^2 \nabla_\alpha\varphi$.

We shall give a possible solution of this consistency problem in the last
Section of the present article.

\section{Transposed-Equi-Affine Theory of Gravity}
Recently a new interesting modification of the ECTG
was proposed by A. Saa in \cite{Saa1}--\cite{Saa5}.
An unexpected solution of the problem with minimal coupling principle
{\em for fields} was discovered.
As a result we have at first a possibility to derive in presence of nonzero
torsion the same equations of motion for fields using MCP both in
action principle and directly in the equations of motion\footnote{A
similar result was reached very recently in different way in
\cite{Pereira1}-\cite{Pereira3}
in the framework of an interesting re-formulation of the standard theory
of gravity in terms of Weitzenb\"ock spaces.}.
It turns out that these equations are equations of autoparallel type
and we reach a new theory of fields in Einstein-Cartan spaces, which
needs to be developed further. Especially, we have to include in this theory
the law of motion of test classical particles and of classical fluids, to
be able to reach results, comparable with experimental evidences. This is
the subject of the present paper.

The main idea of the articles  \cite{Saa1}--\cite{Saa5}
is to make the volume-element $\dV$ compatible
with the affine connection in the four-dimensional Einstein-Cartan space
${\cal M}^{(1,3)}\{g_{\alpha\beta}(x), \Gamma^\alpha_{\beta\gamma}(x)\}$
using the compatibility condition \cite{Saa5}:
\ben
\pounds_v \left(d^DV\!ol\right) = (\nabla_\mu v^\mu) d^DV\!ol,
\la{CC}
\een
$\pounds_v$ being the Lee derivative in direction of the arbitrary
vector field $v^\alpha$, $d^DV\!ol$ being the volume D-form in the space
of dimension $D$. In the case of ECTG $D=1+3=4$, but for a moment we
shall write down the formulae for an arbitrary dimension $D$.
It turns out that the condition (\ref{CC}) is consistent if and only if the
torsion vector $S_{\alpha} = {\sfrac 2 {D-1}}{S_{\alpha \mu}}^\mu$
\footnote{We use the Schouten's normalization conventions \cite{Schouten}
which differs from the original ones in \cite{Saa1}--\cite{Saa5}
and seem to make more apparent some relations.} is potential:
\ben
S_{\alpha} = \nabla_\alpha \Theta \equiv \partial_\alpha \Theta,
\la{GrC}
\een
where $\Theta(x)$ is the corresponding potential.
Then the Saa's condition (\ref{CC}) leads to the form:
\ben
d^DV\!ol = f(x) d^D x =e^{-(D-1)\Theta}\sqrt{|g|} d^D x
\la{Vol}
\een
of the volume element compatible with the affine connection
in Einstein-Cartan space.

Up to the inessential choice of the normalization
of the $\Theta$ field the same volume element was used in \cite{Kl1}
to ensure the hermicity of the quantum hamiltonian in the Schr\"odinger
equation of autoparallel type under condition (\ref{GrC}),
too.\footnote{Essentially the same volume element,
but described in a different form was used to ensure the hermicity
of the quantum hamiltonian for hydrogen
in Kustaanheimo-Stiefel coordinates as early as in \cite{Devreese}.
This procedure for hydrogen leads actually
to a space with torsion, See \cite{Kl1} and the references therein.}

The geometrical and the physical meaning
of the Saa's compatibility condition (\ref{CC}) is not completely clear.
In the original articles \cite{Saa1}--\cite{Saa5} it is commented in
a slightly incorrect way as a condition for covariantly constant volume
under parallel displacement in the affine space. Indeed,
from condition (\ref{CC}) it follows that\,
$\partial_\alpha f - \Gamma^\mu_{\mu\alpha}f=0$ for the scalar density
$f=e^{-(D-1)\Theta}\sqrt{|g|}$,
but this means that it is covariantly constant with respect to
the {\em transposed} connection with coefficients
$(\Gamma^T)^\gamma_{\alpha\beta}=\Gamma^\gamma_{\beta\alpha}$,
i.e. $\nabla^T_\alpha f=0$. In presence of nonzero torsion this
is definitely different from the condition
$\nabla_\alpha f=\partial_\alpha f - \Gamma^\mu_{\alpha\mu}f=0$,
which is fulfilled in the Einstein-Cartan spaces for
the usual volume element with $f_0=\sqrt{|g|}$ \cite{Schouten}.
In other words, in Einstein-Cartan spaces the usual volume element
is covariantly constant due to the metric compatibility of the connection
and there is no need to make any changes to ensure constancy of the volume
under parallel displacement with respect to the basic metric connection.
As we see, one still has to recover the true meaning of the Saa's compatibility
condition (\ref{CC}).

The affine space is called {\em equi-affine}
if the volume element is covariantly constant, i.e. if $\nabla_\alpha f=0$.
But this is not the case for Saa's condition (\ref{CC}), which
is equivalent to the similar relation with respect to
the transposed connection: $\nabla^T_\alpha f=0$.
Therefore we shall call the affine space {\em transposed-equi-affine},
when the condition (\ref{CC}) is fulfilled.
The corresponding theory of gravity will be called
a {\em transposed-equi-affine theory of gravity} (TEATG).

The most important mathematical consequence of the condition (\ref{CC})
is the following generalized Gauss' formula:
\ben
\int_{\cal M}d^DV\!ol \left(\nabla_\mu v^\mu\right) =
\int_{\partial{\cal M}} d^{D-1}\Sigma_\mu v^\mu.
\la{Gauss}
\een
This formula leads to the autoparallel type of equations of motion
for all kind of fields in the space
${\cal M}^{(1,3)}\{g_{\alpha\beta}(x), \Gamma^\alpha_{\beta\gamma}(x)\}$,
derived from the standard action principle for
{\em a nonstandard action integral}:
\ben
{\cal A}_{tot}= {\cal A}_G +  {\cal A}_M =
{\sfrac 1 c}\int\!\dV\,\,{\cal L}_G +
{\sfrac 1 c}\int\!\dV\,\,{\cal L}_M.
\la{Atot}
\een
In accordance with the formula (\ref{Vol}) here
$\dV = e^{-3\Theta}\sqrt{|g|} d^4 x$.
Hence, due to the form of the volume element
in TEATG the lagrangian densities for gravity and for matter are
\ben
\Lambda_G= e^{-3\Theta}\sqrt{|g|}{\cal L}_G, \nonumber\\
\Lambda_M= e^{-3\Theta}\sqrt{|g|}{\cal L}_M.
\la{LD}
\een
Using the standard conventions we write down the lagrangian of gravity:
\ben
{\cal L}_G = - {\sfrac {c^2} {2\kappa}}R =
- {\sfrac {c^2} {2\kappa}} \left(
\stackrel{\{\}}{R} + 6\nabla_\mu S^\mu + 12 S_\mu S^\mu
- {\tilde K}_{\mu\nu\lambda} {\tilde K}^{\mu\nu\lambda}\right),
\la{LG}
\een
$c$ being the velocity of light, $\kappa$ being the Einstein constant.
As usual here $R=g^{\alpha\beta}R_{\alpha\beta}$,
$R_{\alpha\beta}={R_{\mu\alpha\beta}}^\mu$,
${R_{\alpha\beta\mu}}^\nu= 2\left(
\partial_{[\alpha} \Gamma^\nu_{\beta]\mu} +
\Gamma^\nu_{[\alpha|\sigma|}\Gamma^\sigma_{\beta]\mu}\right)$
are the scalar curvature, Ricci tensor and curvature tensor of the affine
connection;\,\,
${\tilde K}_{\mu\nu\lambda}= K_{\mu\nu\lambda} + 2 g_{\mu[\nu} S_{\lambda]}$
is the traceless part of the contorsion:
${\tilde K}^\mu{}_{\mu\nu}={\tilde K}^\mu{}_{\nu\mu}\equiv 0$.
It is connected with the nonzero spin matter
and vanish in vacuum, or in presence only of spin zero matter \cite{Saa1}.

In the present article we will include only spin zero matter
in the lagrangian ${\cal L}_M$.
Therefore we put $\tilde K_{\mu\nu\lambda}\equiv 0$.
This leads to a semi-symmetric affine connection \cite{Schouten}:
\ben
{S_{\alpha\beta}}^\gamma = S_{[\alpha}\delta_{\beta]}^\gamma.
\la{SST}
\een
The basic properties of this special type of affine geometry under additional
condition (\ref{GrC}) are given in the Appendix.

Then using the generalized Gauss' formula (\ref{Gauss}) we obtain
the variations of the action of gravity and action of matter
with respect to metric $g$ and torsion potential $\Theta$:
\ben
\delta_g{\cal A}_G&=& -{\sfrac c {2\kappa}}\int\!\dV\,\,\delta g^{\mu\nu}
\left(G_{\mu\nu} +\nabla_\mu S_\nu -g_{\mu\nu}\nabla_\sigma S^\sigma\right),
\nonumber\\
\delta_g{\cal A}_M&=&{\sfrac 1 {2c}}\int\!\dV\,\,\delta g^{\mu\nu} T_{\mu\nu},
\nonumber\\
\delta_\Theta{\cal A}_G&=& {\sfrac {3c} {2\kappa}}  \int\!\dV\,\,\delta\Theta
\left(R +2\nabla_\sigma S^\sigma\right),
\nonumber\\
\delta_\Theta{\cal A}_M&=&-{\sfrac 3 {c}}  \int\!\dV\,\,\delta\Theta
\left({\cal L}_M -{\sfrac 1 3}{\delta{\cal L}_M \over \delta \Theta}\right),
\la{var}
\een
and the equations of motion for the geometric fields
$g_{\alpha\beta}$ and $\Theta$ in a form:
\ben
G_{\mu\nu} +\nabla_\mu \nabla_\nu \Theta - g_{\mu\nu}\Box\Theta =
{\sfrac \kappa {c^2}}T_{\mu\nu},\nonumber\\
\Box\Theta  = {\sfrac \kappa {c^2}}
\left({\cal L}_M -{\sfrac 1 3}{\delta{\cal L}_M \over \delta \Theta}\right)
-{\sfrac 1 2} R.
\la{GFE}
\een
Here $G_{\mu\nu} = R_{\mu\nu}-{\sfrac 1 2} g_{\mu\nu}$ is the Einstein tensor
for the affine connection, its trace is $G=g^{\mu\nu}G_{\mu\nu}=-R$;
$T_{\mu\nu} = {\lfrac {\delta{\cal L}_M} {\delta g^{\mu\nu}}}$
is the symmetric energy-momentum tensor of the matter, its trace is
$T=g^{\mu\nu}T_{\mu\nu}$; and in accordance with the relation (\ref{GrC})
$\nabla_\sigma S^\sigma=g^{\mu\nu}\nabla_\mu \nabla_\nu \Theta= \Box\Theta$.

Two types of additional relations may be derived from the equations (\ref{GFE}):

1) Algebraic consequences:

\noindent Taking trace of the both sides of the first equation, and combining
the result with the second one we obtain
\ben
\nabla_\sigma S^\sigma= \Box\Theta= -{\sfrac {2\kappa} {c^2}}
\left({\cal L}_M -{\sfrac 1 3}{\delta{\cal L}_M \over \delta \Theta}
+{\sfrac 1 2}T\right)
\la{DivS}
\een
and
\ben
R = {\sfrac {2\kappa} {c^2}}
\left(3{\cal L}_M - {\delta{\cal L}_M \over \delta \Theta} + T\right).
\la{SCurv}
\een

The equation (\ref{DivS}) shows that
under proper boundary conditions in presence only of spinless matter
the torsion is completely determined by its distribution.
Therefore it is convenient to use this equation as an
equation of motion instead of the second equation in the system (\ref{GFE})
\cite{Saa1}.

The equation (\ref{SCurv}) shows that in the case under consideration the
Cartan scalar curvature is completely determined by the matter distribution, too.

2) Differential consequences:

\noindent Calculating the covariant divergence of the
both sides of the first equation and taking into account:

i) the definition of the Einstein tensor $G_{\alpha\beta}$;

ii)the identity $\nabla_{[\alpha}\nabla_{\beta]}v^\gamma\equiv
{\sfrac 1 2}{R_{\alpha\beta\sigma}}^\gamma v^\sigma-
{S_{\alpha\beta}}^\sigma\nabla_\sigma v^\gamma$
which takes place for arbitrary vector field $v^\gamma$;

iii)the identity
$\nabla_\sigma G^\sigma_\alpha + 2 G^\sigma_\alpha S_\sigma = 0$
which follows from the Bianchi identity for semi-symmetric connection:
$\nabla_{[\lambda} {R_{\alpha\beta]\mu}}^\nu =
-2S_{[\lambda}{R_{\alpha\beta]\mu}}^\nu$
-- see Appendix;

\noindent we derive a new important consequence from dynamical equations (\ref{GFE}):
\ben
\nabla_\sigma T_\alpha^\sigma + T^\sigma_\alpha S_\sigma =
{\sfrac {c^2} {2\kappa}}R S_\alpha.
\la{DivT}
\een

This equation gives one more relation between torsion and matter distribution
which is not studied in TEATG up to now.
It is a generalization of the well known local conservation law
${\stackrel{ \{\} } {\nabla} }{}_\sigma T_\alpha^\sigma=0$
for energy-momentum tensor in general relativity, where $S_\alpha \equiv 0$
and may be rewritten in a form
\ben
{\stackrel{ \{\} } {\nabla} }{}_\sigma T_\alpha^\sigma =
3 \left( T^\sigma_\alpha
+ \left({\cal L}_M - {\sfrac 1 3}{\delta{\cal L}_M \over \delta \Theta}\right)
\delta^\sigma_\alpha \right)S_\sigma.
\la{Div{}T}
\een

To have a complete set of dynamical equations one has to add
to the above relations the equations of motion of the very matter.
This will be done in the next sections by proper choice of the
matter lagrangian and of the corresponding variational principle.

\section{Scalar field in TEATG}
Consider the scalar field $\phi(x)$ with lagrangian
\ben
{\cal L}_\phi =
{\sfrac 1 2} g^{\mu \nu}(x) \nabla_\mu \phi(x) \nabla_\nu \phi(x)
-{\sfrac 1 2} m_\phi^2 \phi^2(x) - V(\phi(x)).
\la{SFL}
\een
Putting ${\cal L}_M = {\cal L}_\phi $ in the action (\ref{Atot})
and using usual variational principle based on
the generalized Gauss' formula (\ref{Gauss}) we obtain
the autoparallel type of field equation
\ben
\Box \phi + m_\phi^2 \phi + V^\prime(\phi) = 0.
\la{SFE}
\een
The energy-momentum tensor and its trace are:
\ben
T_{\alpha\beta}=\nabla_\alpha \phi \nabla_\beta \phi -
g_{\alpha\beta}{\cal L}_\phi, \,\,\,
T = g^{\mu \nu} \nabla_\mu \phi \nabla_\nu \phi - 4{\cal L}_\phi.
\la{E-Msf}
\een
Then ${\cal L}_M -{\sfrac 1 3}{\delta{\cal L}_M \over \delta \Theta}
+{\sfrac 1 2}T = {\sfrac 1 2} m_\phi^2 \phi^2 + V(\phi)$ and
$3{\cal L}_M -{\delta{\cal L}_M \over \delta \Theta} +T =
{\sfrac 1 2} g^{\mu \nu} \nabla_\mu \phi \nabla_\nu \phi
+{\sfrac 1 2} m_\phi^2 \phi^2 + V(\phi)$. Hence,
\ben
\Box \Theta = -{\sfrac {2\kappa} {c^2}}
\left({\sfrac 1 2} m_\phi^2 \phi^2 + V(\phi)\right) ,
\la{Theta-phi}
\een
\ben
R = {\sfrac {2\kappa} {c^2}}
\left( {\sfrac 1 2} g^{\mu \nu} \nabla_\mu \phi \nabla_\nu \phi
+{\sfrac 1 2} m_\phi^2 \phi^2 + V(\phi)\right).
\la{R-phi}
\een
As a consequence we obtain the universal relation
\ben
\Box \Theta + R = {\sfrac \kappa {c^2}}
g^{\mu \nu} \nabla_\mu \phi \nabla_\nu \phi
\la{The-R-phi}
\een
which does not depend on the mass $m_\phi = const$
and on the self-interaction $V(\phi)$ of the scalar field $\phi$.
The equation of motion (\ref{SFE}) gives
\ben
\nabla_\sigma T_\alpha^\sigma &=&
2 {S_\alpha}^{\mu\nu}\nabla_\mu \phi \nabla_\nu \phi=
\biggl(g^{\mu \nu} \nabla_\mu \phi \nabla_\nu \phi \, \delta^\sigma_\alpha -
\nabla_\alpha \phi \nabla^\sigma \phi \biggr) S_\sigma
, \,\,\,\,\hbox{or}
\nonumber\\
{\stackrel{ \{\} } {\nabla} }{}_\sigma T_\alpha^\sigma &=&
3 S_\sigma \nabla^\sigma \phi \nabla_\alpha \phi.
\la{EMDiv}
\een
Substitution of this result into relation (\ref{DivT}) and the use of
the equation (\ref{R-phi}) shows that in the case of scalar field
with arbitrary mass $m_\phi$ and arbitrary self-interaction $V(\phi)$
the equation (\ref{DivT}) is identically fulfilled in TEATG.

One may turn back the last result:
the equation of motion (\ref{SFE}) of the scalar field
with an energy-momentum tensor (\ref{E-Msf}) may be derived from relation
(\ref{DivT}) which follows from the Bianchi identity (\ref{Bianchi})
in TEATG just in the same way as in the general relativity.
In other words, the nonlinear field equations
for geometric fields (\ref{GFE}) via the Bianchi identity (\ref{Bianchi})
imply the equation of motion (\ref{SFE}) of the scalar field defined
by the energy-momentum tensor (\ref{E-Msf}). Hence, the matter field equation
(\ref{SFE}) may be considered as a compatibility condition of the equations
(\ref{GFE}) for the geometric fields $g_{\alpha\beta}$ and $\Theta$.

\section{Relativistic Fluid in TEATG}
The change of such a basic notion as the volume element in the space
${\cal M}^{(1,3)}\{g_{\alpha\beta}(x), \Gamma^\alpha_{\beta\gamma}(x)\}$,
as suggested in \cite{Saa1}-\cite{Saa5}, requires a very careful analysis.

In the framework of TEATG we develop the relativistic fluid's theory
using both  Euler variables $x=\{x^\alpha\}$ of the local frame,
and  Lagrange  variables $\vec{r}$ of the co-moving frame
following the reference \cite{Fock}
(but in the slightly different notations of the reference \cite{LL}
which are accepted throughout the present article).
We denote by $u^\alpha(x)$ the velocity field, normalized according to
the equation $g_{\mu\nu}u^\mu u^\nu=1$.
Let $x^\alpha(t,\vec{r})$ denote the trajectory of a fluid's particle
in Lagrange variables. Then the relation
$\dot x^\alpha = \sqrt{g_{\mu\nu}\dot x^\mu \dot x^\nu}\, u^\alpha(x)$
when considered as a system of ordinary differential equations for
$x^\alpha(t,\vec{r})$ under initial conditions
$t_{in}=0, \{x^{\alpha=1,2,3}_{in}\} =\vec{r}$, according to Liouville theorem
imply the equality:
\ben
\partial_\alpha\left( J^{-1}
{\sqrt{g_{\mu\nu}\dot x^\mu \dot x^\nu}}\, u^\alpha\right)=0,
\la{Liouville}
\een
where $J = {\frac {D(x(t,\vec{r}))} {D(t,\vec{r})}}$ is the jacobian
of the transition from Euler to Lagrange  variables.
The existence of this result reflects only the structure of the space
${\cal M}^{(4)} \ni \{x^\alpha\}$ as a differentiable manifold
and does not depend on its metric, or on the affine connection\footnote{The
metric tensor enters into equation (\ref{Liouville}) "incidentally"
because of the normalization of the four-velocity field $u^\alpha(x)$.}.
In this sense it presents an universal relation.

The lagrangian of the fluid with internal pressure $p$ will be taken in
standard form:
\ben
{\cal L}_\mu = - \varepsilon = -\mu c^2 - \mu\Pi,
\la{FLagrangian}
\een
$\mu(x)$ being properly defined fluid's density in Euler variables,
$\Pi$ being the elastic potential energy of the fluid:
$\db \Pi = - p d({\frac 1 \mu})$,
where the symbol "$\db$\,'' denote a differential form, which is not exact.

The main problem of the present theory of the relativistic fluid is
the choice of the continuity condition,
which together with the lagrangian (\ref{FLagrangian})
actually defines what we mean by "fluid'',
as well as the choice of the variational principle
for fluid's particles' trajectories.
As we shall see, there exist different possibilities and
at present we are not able to reach unambiguously a theory of particles in
the TEATG. Moreover, due to the choice of the fluid's definition a different
interpretations of the very theory are possible.

The continuity condition describes the conservation of the fluid's matter.
In its four dimensional form it reads
\ben
\int_{\partial \Delta^{(1,3)}} d^3\Sigma_\alpha \,\mu(x) u^\alpha(x) = 0,
\la{IntCE}
\een
where $d^3\Sigma_\alpha$ is a proper three-dimensional surface element,
which depends on the choice of the four-dimensional volume element
$\dV$ via the Gauss' formula (\ref{Gauss}),
and $\Delta^{(1,3)} \in
{\cal M}^{(1,3)}\{g_{\alpha\beta}(x), \Gamma^\alpha_{\beta\gamma}(x)\}$ is
an arbitrary domain.
The relation (\ref{IntCE}) shows that the continuity condition
actually does not depend on the very metric and connection of the space
${\cal M}^{(1,3)}\{g_{\alpha\beta}(x), \Gamma^\alpha_{\beta\gamma}(x)\}$,
but just on the choice of the volume element
\footnote{The relation (\ref{IntCE}) may be rewritten in the language of
differential forms using Hodge star operator
which itself depends just on the choice of the volume element \cite{Saa5}.}.

\subsection{Relativistic Fluid in a Strict Saa's Model}
If we take seriously the volume element (\ref{Vol}) as an universal volume
element in TEATG, we must use it in the continuity condition, too.
Then according to generalized Gauss' formula (\ref{Gauss}) we can
rewrite the relation (\ref{IntCE}) as
$\int_{\Delta^{(1,3)}}\dV\, \nabla_\alpha \left( \mu(x) u^\alpha(x)\right) = 0$,
or in a form the following continuity equation of an {\em autoparallel} type:
\ben
\nabla_\alpha \left( \mu(x) u^\alpha(x) \right) = 0.
\la{ACE}
\een

Now, the comparison of the universal relation (\ref{Liouville})
with the equation (\ref{ACE}), written in a form
$\left(e^{-3\theta}\sqrt{|g|}\right)^{-1}\partial_\alpha
\left(e^{-3\theta}\sqrt{|g|}\mu(x) u^\alpha(x)\right)= 0$,
brings us to the following explicit expression for
the fluid's density in  Lagrange variables:
\ben
\mu = \mu_0(\vec{r})
\left( J e^{-3\theta} \sqrt{|g|} \right)^{-1}
\sqrt{g_{\mu\nu}\dot x^\mu \dot x^\nu}.
\la{AFD}
\een
Here $\mu_0(\vec{r})$ is the fluid's density in the co-moving system
(i.e. $\mu_0(\vec{r})$ is the analog of the rest mass of the particles).

As a consequence we obtain the action for single particle from the
fluid's action ${\cal A}_\mu = \int\dV\, {\cal L}_\mu$ putting into formula
(\ref{FLagrangian}) $\Pi=0$ and into equation (\ref{AFD})
$\mu_0(\vec{r})= m_0 \delta(\vec{r})$. This way we reach the
action integral (\ref{spA}) which  does not feel the torsion of the space
${\cal M}^{(1,3)}\{g_{\alpha\beta}(x), \Gamma^\alpha_{\beta\gamma}(x)\}$
in contrast to the case of action integrals for fields in TEATG.
Actually the action integral for dust matter ($\Pi=0$) does not feel the
torsion, too, due to the relation (\ref{AFD}) which follows
the continuity equation (\ref{ACE}) based on the volume element (\ref{Vol}).

Then, using equations
(\ref{Atot}), (\ref{var}), (\ref{FLagrangian}), (\ref{AFD})
and the procedure described in \cite{Fock},
we reach the usual form of the fluid's energy-momentum tensor
\cite{Fock}, \cite{LL}:
\ben
T^{\alpha\beta} &=& (\varepsilon + p)u^\alpha u^\beta - p\,g^{\alpha\beta},
\nonumber\\
T &=& \varepsilon - 3 p.
\la{FEMT}
\een

After an additional calculation which gives
${\cal L}_\mu -{\sfrac 1 3}{\delta{\cal L}_\mu \over \delta \Theta} = p$
we are ready to write down explicitly the equations of motion (\ref{GFE})
of geometric fields $g_{\alpha\beta}$ and $\Theta$ and the additional
relations (\ref{SCurv}), (\ref{DivT}), (\ref{Div{}T})  in presence of fluid:
\ben
G_{\mu\nu} + (\nabla_\mu \nabla_\nu - g_{\mu\nu}\Box)\Theta &=&
{\sfrac \kappa {c^2}}\biggl((\varepsilon + p)u^\mu u^\nu
- p\,g^{\mu\nu}\biggr),\nonumber\\
\Box\Theta  &=& - {\sfrac \kappa {c^2}} (\varepsilon - p);
\la{GFE_Flu}
\een
\ben
R = {\sfrac {2\kappa} {c^2}}\varepsilon,
\la{SCurv_Flu}
\een
\ben
\nabla_\sigma T_\alpha^\sigma &=&
(\varepsilon + p)\left(\delta^\sigma_\alpha-u^\sigma u_\alpha\right)S_\sigma,
\,\,\,\,\,\,\,\,\,\,\,\,\hbox{or}\nonumber\\
{\stackrel{ \{\} } {\nabla} }{}_\sigma T_\alpha^\sigma &=&
3(\varepsilon + p)\,u^\sigma u_\alpha S_\sigma.
\la{DivT_Flu}
\een

To derive the fluid's equation of motion we need to calculate the variation
$\delta_x {\cal A}_\mu = \int\dV\,\, \delta_x {\cal L}_\mu $ under variation
of the trajectories of fluid's particles. The key step in this direction is
the calculation of the variation $\delta_x \mu$, which may be represented
according to the equality (\ref{AFD}) in a form:
\ben
\delta_x \mu = \delta_x\mu_1 + \delta_x\mu_2 =
-\mu {\sfrac  {\delta_x\left( J e^{-3\theta} \sqrt{|g|} \right)}
{\left( J e^{-3\theta} \sqrt{|g|} \right)}}
+ \mu {\sfrac {\delta_x \left(\sqrt{g_{\mu\nu}\dot x^\mu \dot x^\nu}\right)}
{\sqrt{g_{\mu\nu}\dot x^\mu \dot x^\nu}}}.
\la{muVar}
\een

We obtain for the first term
$\delta_x\mu_1 = -\nabla_\alpha\left(\mu \delta x^\alpha\right)$
applying the corresponding procedure of the reference \cite{Fock} to the formula
(\ref{AFD}). This result is determined completely by the choice of the
volume element (\ref{Vol}) in the continuity condition (\ref{IntCE}).

The second term $\delta_x\mu_2 =
\mu {\sfrac {\delta_x \left(\sqrt{g_{\mu\nu}\dot x^\mu \dot x^\nu}\right)}
{\sqrt{g_{\mu\nu}\dot x^\mu \dot x^\nu}}}$
requires a justification of the variational principle for particle trajectories.
If we accept the usual variational principle with fixed boundary
for particle trajectories, as we did for fields according to the articles
\cite{Saa1}-\cite{Saa5}, we will have
\ben
\delta_x {\frac d {dt}} - {\frac d {dt}} \delta_x = 0.
\la{ComRel}
\een
Then we obtain a geodesic motion for a single free test particle
according to the equation (\ref{spEG}) with torsion force
${\cal F}_\alpha =
m c^2 (\delta_\alpha^\beta - u_\alpha u^\beta)\nabla_\beta \Theta$
and
\ben
\delta_x \mu &=&
-\nabla_\alpha\left(\mu \delta x^\alpha\right) + \mu u_\alpha u^\beta
{\stackrel{{\{\}}}{\nabla}}_\beta \left(\delta x^\alpha\right)
\nonumber \\
&=& -\nabla_\alpha \biggl( \mu
 \left(\delta_\beta^\alpha - u^\alpha u_\beta \right)\delta x^\beta \biggr)
- \left(\mu u^\beta {\stackrel{{\{\}}}{\nabla}}_\beta u_\alpha\right)
\delta x^\alpha.
\la{muVar_}
\een
Now we calculate the variation $\delta_x {\cal L}_\mu = -\delta_x \varepsilon$
of the lagrangian (\ref{FLagrangian}).
A straightforward generalization of the procedure, described for this purpose
in \cite{Fock} in which one must take into account the new relation
(\ref{muVar_}) results to the formula
\ben
\delta_x {\cal L}_\mu &=& \nabla_\alpha \biggl( (\varepsilon + p)
 \left(\delta_\beta^\alpha - u^\alpha u_\beta \right)\delta x^\beta \biggr)
\nonumber \\
&+& \biggl( (\varepsilon + p)u^\beta {\stackrel{{\{\}}}{\nabla}}_\beta u_\alpha
-\left(\delta^\beta_\alpha - u_\alpha u^\beta \right)\partial_\beta p\biggr)
\delta x^\alpha.
\la{Gdelta_xL_mu}
\een
Hence, using the standard variational principle $\delta_x {\cal A}_{tot}=0$,
based on the generalized Gauss' formula (\ref{Gauss}), one reaches
the following equation of motion of {\em geodesic} type for the fluid:
\ben
(\varepsilon + p) u^\beta {\stackrel{{\{\}}}{\nabla}}_\beta u_\alpha =
\left(\delta^\beta_\alpha - u_\alpha u^\beta \right)
{\stackrel{{\{\}}}{\nabla}}_\beta p.
\la{GFluEM}
\een
Making use of the formula (\ref{GrC}) we can write down this equation
in a form
\ben
(\varepsilon + p) u^\beta \nabla_\beta u_\alpha =
\left(\delta^\beta_\alpha - u_\alpha u^\beta \right)
\biggl(\nabla_\beta p + (\varepsilon + p)\nabla_\beta\Theta\biggr).
\la{GFluEM_}
\een
It is not hard to check that the additional condition (\ref{DivT_Flu})
follows from the equations (\ref{FEMT}) and (\ref{GFluEM_}).
We may convert this statement and derive the fluid's equation of motion
(\ref{GFluEM_}) from the equation (\ref{FEMT}) as a definition
of the energy-momentum tensor and from the relation (\ref{DivT_Flu}),
which follows the Bianchi identity just as in the general relativity.

We see that the Saa's program to comply the use of the MCP in the
equations of motion with the use of the MCP in the action principle
introducing a new universal volume element (\ref{Vol}) fails
in the case of relativistic fluid.
If we use the new volume element in the continuity condition,
we get continuity equation (\ref{ACE}) of autoparallel type,
but from a standard variational principle with the same
volume element in action integral we obtain the geodesic type equation
of motion (\ref{GFluEM_}) with torsion force density\,
${\cal F}_\alpha= (\varepsilon + p)
\left(\delta^\beta_\alpha - u_\alpha u^\beta \right) \nabla_\beta\Theta$.

Hence, the Saa's program turns to be not self-consistent in its original form.
Different modifications of this program may be suggested.

\subsection{Modification of the Variational Principle for Particles}
We may try to overcome the above described problem modifying the variational
principle for particles in presence of torsion.
Following the reference \cite{Timan}, we can {\em postulate} instead of
the commutation relation (\ref{ComRel}) the new one\footnote{Recently
in \cite{Kl2}, \cite{Kl4} it was proposed once more to postulate
the commutation relation (\ref{ComRel_T}).}:
\ben
\biggl( \delta_x {\frac d {dt}} - {\frac d {dt}} \delta_x \biggr) x^\alpha=
2 {S_{\mu\nu}}^\alpha \dot x^\mu \delta x^\nu.
\la{ComRel_T}
\een

The same commutation relation was {\em derived}
in presence of nonzero torsion in a teleparallel Weitzenb\"ock space
${\cal M}^{(1,3)}\{g_{\alpha\beta}(x), \Gamma^\alpha_{\beta\gamma}(x)\}$
(i.e. when Cartan curvature tensor vanish) in \cite{F3}, \cite{F4}
and used in the case of relativistic particles in \cite{F1} to derive
the autoparallel type equations of motion (\ref{spA}) in this case.
Unfortunately,
in the general case with nonzero torsion and nonzero Cartan curvature
is not clear up to now how to prove that the relation (\ref{ComRel_T})
take place, or that it must be replaced with some more general one.
Therefore in the present article we briefly outline only some possible
consequences of this modification of variational principle for particles
trajectories.

The basic new result which follows from equation (\ref{ComRel_T}) is that now
we obtain for the term $\delta_x\mu_2$ in formula (\ref{muVar})
$\delta_x\mu_2 = \mu u_\alpha u^\beta \nabla_\beta \left(\delta x^\alpha\right)$.
Hence, now
\ben
\delta_x {\cal L}_\mu &=& \nabla_\alpha \biggl( (\varepsilon + p)
 \left(\delta_\beta^\alpha - u^\alpha u_\beta \right)\delta x^\beta \biggr)
\nonumber \\
&+& \biggl( (\varepsilon + p)u^\beta \nabla_\beta u_\alpha
-\left(\delta^\beta_\alpha - u_\alpha u^\beta \right)\partial_\beta p\biggr)
\delta x^\alpha
\la{Adelta_xL_mu}
\een
and the modified variational principle for fluid's particles after all
brings us to the equation of motion of an {\em autoparallel} type:
\ben
(\varepsilon + p) u^\beta \nabla_\beta u_\alpha =
\left(\delta^\beta_\alpha - u_\alpha u^\beta \right)\nabla_\beta p.
\la{AFluEM}
\een
Together with the continuity equation (\ref{ACE}) the new equation of motion
(\ref{AFluEM}) leads to the following local conservation law for the fluid's
energy-momentum tensor (\ref{FEMT}):
\ben
\nabla_\sigma T_\alpha^\sigma=0,
\la{AFEMC}
\een
which looks like a natural generalization of the corresponding local
conservation law in the general relativity.

But in contrast to the previous case of geodesic motion of the fluid
the autoparallel equation (\ref{AFluEM}) is compatible with the
identity (\ref{DivT_Flu}) only if the additional requirement
$(\delta^\sigma_\alpha-u^\sigma u_\alpha) S_\sigma=0$ is fulfilled,
i.e. if the torsion vector is parallel to the velocity of the fluid:
\ben
S_\alpha = -\sigma u_\alpha,
\la{u_S_longit}
\een
$\sigma(x)$ being some new scalar field to be determined. Then the condition
(\ref{GrC}) permits us to rewrite the relation (\ref{u_S_longit}) in a form
$d\Theta = - \sigma u_\alpha dx^\alpha$
which according to the Frobenius theorem yields the restriction
$\epsilon^{\alpha\mu\nu\lambda} u_\mu \partial_\nu u_\lambda \equiv 0$
for the four-velocity $u_\alpha$ of the fluid.

The equation of motion of the field $\Theta$ in the system (\ref{GFE_Flu})
together with continuity equation (\ref{ACE}) gives an equation for the field
$\sigma$:
$ u^\alpha \nabla_\alpha({\frac \sigma \mu}) =
\kappa{\frac {\varepsilon -p} {\mu c^2}}$.
It may be solved in  Lagrange  variables in a form
$\sigma =
\kappa \mu \int {\frac {\varepsilon -p} {\mu c^2}}
\sqrt{g_{\mu\nu}\dot x^\mu \dot x^\nu}dt$.
In the case of a dust matter this gives
$\sigma = \kappa \mu  \int\sqrt{g_{\mu\nu}\dot x^\mu \dot x^\nu}dt$
and shows that in the variant of theory under consideration
the point particle will looks like
a specific space-time (autoparallel) line defect with torsion vector
defined by relation (\ref{u_S_longit})
which is similar but not identical to space-time dislocation \cite{Kl5}.

\subsection{Theory with Usual Volume Element in the Continuity Condition for
Fluid}
Another type of modification of the Saa's original idea may be reached if we
accept to use the usual volume element $dV_0= d^4x \sqrt{|g|}$
in the integral form of the continuity condition (\ref{IntCE})
and the Saa's modified volume element (\ref{Vol})
in the action integrals (\ref{Atot}).

The same consideration as in Section 4.1 now gives
the continuity equation
\ben
\nabla^T_\alpha \left( \mu(x) u^\alpha(x) \right) \equiv
{\stackrel{{\{\}}}{\nabla}}_\alpha \left( \mu(x) u^\alpha(x) \right) = 0.
\la{GCE}
\een
It may be interpreted both as an equation of autoparallel type
with respect to the transposed affine connection, or as a geodesic type
of relation,
and for the fluid's density imply the equality
\ben
\mu = \mu_0(\vec{r}) \left( J \sqrt{|g|} \right)^{-1}
\sqrt{g_{\mu\nu}\dot x^\mu \dot x^\nu}
\la{GFD}
\een
instead of the relation (\ref{AFD}).

Putting this expression into the standard fluid's lagrangian (\ref{FLagrangian})
we derive the same form (\ref{FEMT}) of the fluid's energy-momentum tensor
because the new fluid's density (\ref{GFD}) depends on the metric
$g_{\alpha\beta}$ in the same way as the density (\ref{AFD})\footnote{
In the variant of the theory developed in the present
section we would have to replace the Saa's volume element $\dV$
with the usual one $\dV_0$ in the second of the equations (\ref{var})
which defines the energy-momentum tensor $T_{\mu\nu}$, too.
Then the factor $e^{-3\Theta}$ would be absorbed in it
and the formulae with $T_{\mu\nu}$ would be changed correspondingly.
Here we prefer to keep as far as possible an uniform treatment of all variants
of the TEATG and therefore we use without changes
the second of the equations (\ref{var}) as a definition
of the energy-momentum tensor $T_{\mu\nu}$.}.
But in contrast to (\ref{AFD}) the fluid's density (\ref{GFD})
does not depend on the torsion potential $\Theta$ and this yields a new
fluid's dynamics. As a result the very fluid's lagrangian  (\ref{FLagrangian})
becomes independent of torsion potential and the whole dependence of the
fluid's action on it goes into the factor $e^{-3\theta}$ in the
Lagrange density $\Lambda_\mu= e^{-3\Theta}\sqrt{|g|}{\cal L}_\mu$
just as in the case of other fields in formulae (\ref{LD}).
In particular, we get for the single particle an action
\ben
{\cal A}[x(t)]=
-mc\int e^{-3\theta}\sqrt{g_{\mu \nu}(x(t)) \dot x^\mu(t) \dot x^\nu(t)}\,dt=
-mc\int e^{-3\theta}ds
\la{New_spA}
\een
instead of the action (\ref{spA}).

The fluid's density (\ref{GFD}) is precisely the same as in
the general relativity.
Hence, if we accept the standard variational principle
based on the commutation relation (\ref{ComRel}) for particles,
we can share without changes the result of the reference \cite{Fock}
for the variation of the fluid's lagrangian
and rewrite it in terms of the affine connection:
\ben
\delta_x {\cal L}_\mu = \hskip 6.truecm \nonumber \\
{\stackrel{{\{\}}}{\nabla}}_\alpha \biggl( (\varepsilon + p)
\left(\delta_\beta^\alpha - u^\alpha u_\beta \right)\delta x^\beta \biggr)
+ \biggl( (\varepsilon + p)u^\beta {\stackrel{{\{\}}}{\nabla}}_\beta u_\alpha
-\left(\delta^\beta_\alpha - u_\alpha u^\beta \right)\partial_\beta p\biggr)
\delta x^\alpha =
\nonumber \\
\nabla_\alpha \biggl( (\varepsilon + p)
\left(\delta^\beta_\alpha - u_\alpha u^\beta \right)\delta x^\beta \biggr) +
\hskip 7.8truecm \nonumber \\
\biggl( (\varepsilon + p)u^\beta \nabla_\beta u_\alpha
-\left(\delta^\beta_\alpha - u_\alpha u^\beta \right)
\left(\vbox to 12pt {} \nabla_\beta p -
2\,(\varepsilon + p)\nabla_\beta \Theta\right)\biggr)
\delta x^\alpha. \hskip 1.truecm
\la{delta_xL_mu}
\een
This yields the fluid's equation of motion
\ben
(\varepsilon + p)u^\beta \nabla_\beta u_\alpha =
\left(\delta^\beta_\alpha - u_\alpha u^\beta \right)
\nabla_\beta p + {\cal F}_\alpha
\la{New_FEM}
\een
which is not of autoparallel type, nor of geodesic one and includes the
torsion force density
${\cal F}_\alpha = - 2\,(\varepsilon + p)
\left(\delta^\beta_\alpha - u_\alpha u^\beta \right)
\nabla_\beta \Theta$.\footnote{We shall not write down the fluid's
equation of motion which one reaches if one uses once more the modified
variational principle with the commutation relation (\ref{ComRel_T}).
In this case the first covariant derivative in the equation (\ref{delta_xL_mu})
will be ${\stackrel{{\{\}}}{\nabla}}_\alpha$, and the second one --
$\nabla_\alpha$. As a result the corresponding equation
will have the form (\ref{New_FEM})
but the torsion force will have a coefficient $3=D-1$ instead of the
coefficient $2 = D-2$. Hence, the four variants of fluid's theory
in TEATG, described in present article yield the torsion force
${\cal F}_\alpha = - q\,(\varepsilon + p)
\left(\delta^\beta_\alpha - u_\alpha u^\beta \right)\nabla_\beta \Theta$
with q= -1,0,2,3 in the fluid's equation (\ref{New_FEM}).}

Now we have
${\cal L}_\mu -{\sfrac 1 3}{\delta{\cal L}_\mu \over \delta \Theta} =
-\varepsilon$. Then the equations of motion (\ref{GFE})
of the geometric fields $g_{\alpha\beta}$ and $\Theta$ and the additional
relations (\ref{SCurv}), (\ref{DivT}), (\ref{Div{}T})  in presence of fluid
with density (\ref{GFD}) are:
\ben
G_{\mu\nu} + (\nabla_\mu \nabla_\nu - g_{\mu\nu}\Box)\Theta &=&
{\sfrac \kappa {c^2}}\biggl((\varepsilon + p)u_\mu u_\nu
- p\,g_{\mu\nu}\biggr),\nonumber\\
\Box\Theta  &=&  {\sfrac \kappa {c^2}} (\varepsilon + 3 p);
\la{New_GFE_Flu}
\een
\ben
R = - {\sfrac {2\kappa} {c^2}}(2 \varepsilon + 3 p),
\la{New_SCurv_Flu}
\een
\ben
\nabla_\sigma T_\alpha^\sigma &=&
-(\varepsilon+p)\left( 2 \delta^\sigma_\alpha+u^\sigma u_\alpha\right)S_\sigma,
\,\,\,\,\,\,\,\,\,\,\,\,\hbox{or}\nonumber\\
{\stackrel{ \{\} } {\nabla} }{}_\sigma T_\alpha^\sigma &=&
-3(\varepsilon + p)\left(\delta^\beta_\alpha - u_\alpha u^\beta\right) S_\beta.
\la{New_DivT_Flu}
\een

The equation of motion (\ref{New_FEM})
is compatible with the identity (\ref{New_DivT_Flu})
without any additional restrictions and may be derived using this identity,
the dynamical equations (\ref{New_GFE_Flu}) of the geometric fields,
the form of the energy-momentum tensor (\ref{FEMT}),
and the continuity equation (\ref{GCE}).

One must confess that this modification of the Saa's program
is not completely successful,
due to the torsion force ${\cal F}_\alpha$ in the equation (\ref{New_FEM}).
Nevertheless it is quite curious,
because it leads to a new physical interpretation of the TEATG,
as we shall see in the Section 6 of present article.

\section{Local Energy-Momentum Conservation}
In the previous Sections we saw that in TEATG
both the absolute divergence of the matter energy-momentum tensor with
respect to the basic affine connection
and its absolute divergence with respect to the Levi-Cevita one
do not vanish in general --
See equations (\ref{EMDiv}), (\ref{DivT_Flu}), and (\ref{New_DivT_Flu}).
The only exception was the second variant of fluid's theory --
See equation (\ref{AFEMC}),
where an additional restriction (\ref{u_S_longit})
on the torsion vector occurs.

It is well known that in the general relativity still exist some
difficulties with the conservation of the energy-momentum.
At present stage of affairs the corresponding situation in TEATG is worst --
we have no even a local conservation of these fundamental physical
quantities.
But there is a chance to have at least a local conservation law

1) of autoparallel type:
\ben
\nabla_\sigma T_\alpha^\sigma = 0
\la{AConserv_EM}
\een
if in addition we superimpose the condition
\ben
T^\sigma_\alpha S_\sigma = {\sfrac {c^2} {2\kappa}}R S_\alpha
\la{AACT}
\een
for nontrivial solutions of the torsion field equation; or

2) of geodesic type:
\ben
{\stackrel{ \{\} } {\nabla} }{}_\sigma T_\alpha^\sigma =0
\la{GConserv_EM}
\een
for nontrivial solutions if in addition we superimpose the condition
\ben
T^\sigma_\alpha S_\sigma =
-\left({\cal L}_M - {\sfrac 1 3}{\delta{\cal L}_M \over \delta \Theta}\right)
S_\alpha.
\la{GACT}
\een
These additional requirements mean that to have a local
conservation law similar to that one in general relativity
the torsion vector must be an eigenvector of the matter's
energy-momentum tensor with eigenvalue
${\sfrac {c^2} {2\kappa}}R=
\left(3{\cal L}_M - {\delta{\cal L}_M \over \delta \Theta} + T\right)$, or
$(-1)\left({\cal L}_M -
{\sfrac 1 3}{\delta{\cal L}_M \over \delta \Theta}\right)$
respectively.

The local conservation laws (\ref{AConserv_EM}) and (\ref{GConserv_EM})
lead to different consequences.
A priori it is not obvious which one of them to chose in TEATG, if any.
There exists a possibility to superimpose some other additional restriction
on the torsion vector, too.
If we require no additional conditions like (\ref{AACT}), or (\ref{GACT}),
developing the theory only on the bases of the equation (\ref{DivT}),
this would mean that we have to look for a new conservation law of the
energy-momentum of the whole system,
including a properly defined energy-momentum of the
geometric fields $g_{\alpha\beta}$ and $\Theta$,
i.e. we will be forced to
associate with these fields some new {\em physical} degrees of freedom which
carry a part of the energy-momentum of the whole system of matter and
geometric fields.
For a sensible decision of this problem a further development
of the physical content of the theory is needed.
We could make the right choice between different alternatives only after
considering the corresponding consequences for some specific physical problems.
In addition the compatibility of the accepted constrains (if any)
with the previous equations must be investigated.
In the present article we give only some preliminary notes
on the last problems.

\subsubsection{The Case of Scalar field}
It turns out that the scalar field energy-momentum tensor (\ref{E-Msf})
has got two different eigenvalues, precisely these we need:
1) the eigenvalue ${\sfrac {c^2} {2\kappa}}R$ --
with eigenvector $\nabla_\alpha\phi=\partial_\alpha\phi$; and
2) the eigenvalue $(-1)\left({\cal L}_\phi -
{\sfrac 1 3}{\delta{\cal L}_\phi \over \delta \Theta}\right)$
-- with eigenvector $t_\alpha$ -- an arbitrary vector which is orthogonal to
$\nabla_\alpha\phi$:\, $g^{\alpha\beta}t_\alpha \nabla_\beta\phi=0$.
Hence, for scalar field both additional conditions
(\ref{AACT}) and (\ref{GACT}) are possible. Then:

a) In the case of the condition (\ref{AACT}) we will have
$S_\alpha = \partial_\alpha \Theta = - \sigma \partial_\alpha  \phi$,
i.e. the torsion vector must be longitudinal with respect
to $\nabla_\alpha\phi$, which imply $\Theta = \Theta(\phi)$.

b) In the case of the condition (\ref{GACT}) we will have
$g^{\alpha\beta}S_\alpha \nabla_\beta\phi=
g^{\alpha\beta}\nabla_\alpha \Theta \nabla_\beta\phi=0$, i.e. the torsion
vector must be transversal with respect to $\nabla_\alpha\phi$.

In both cases the dynamics simplifies significantly and seems to be
compatible with the additional conditions, but this needs further
study.

\subsubsection{The Case of Spinless fluid}
The energy-momentum tensor (\ref{FEMT}) for the fluid in all cases of fluid's
dynamics has two different eigenvalues, too, precisely:
1) the eigenvalue $\varepsilon$ -- with eigenvector $u_\alpha$; and
2) the eigenvalue $(-p)$ -- with eigenvector $t_\alpha$ --
an arbitrary vector which is orthogonal to
$\nabla_\alpha\phi$:\, $g^{\alpha\beta}t_\alpha \nabla_\beta\phi=0$.
Next consideration depends on the variant of fluid's dynamics we accept:

I. In the case of fluid's dynamics described in Section 4.1  we have
${\sfrac {c^2} {2\kappa}}R= \varepsilon$,
$(-1)\left({\cal L}_\mu -
{\sfrac 1 3}{\delta{\cal L}_\mu  \over \delta \Theta}\right)= -p$.
Hence, we may superimpose each of the conditions (\ref{AACT}), or (\ref{GACT}). Then:

a) In the case of the condition (\ref{AACT}) we will have
$S_\alpha = \partial_\alpha \Theta = - \sigma u_\alpha$,
i.e. the torsion vector must be longitudinal with respect
to the four-velocity $u_\alpha$ of the fluid.

b) In the case of the condition (\ref{GACT}) we will have
$g^{\alpha\beta}S_\alpha u_\beta=0$, i.e. the torsion vector must be
transversal with respect to the four-velocity $u_\alpha$ of the fluid.

II. In the case of fluid's dynamics described in Section 4.2 the
modified action principle for fluid's particles yields
the condition (\ref{AACT}). Then it is the only possible additional condition
and leads to the torsion vector which is longitudinal with respect
to the four-velocity $u_\alpha$ of the fluid.

III. In the case of fluid's dynamics described in Section 4.3 we have
${\sfrac {c^2} {2\kappa}}R= 2\varepsilon +3p$ which {\em is not} an eigenvalue
of the energy-momentum tensor, and the only possibility to superimpose an
additional condition of the discussed kind is to use the eigenvector $u_\alpha$
which this time corresponds to the eigenvalue $(-1)\left({\cal L}_\mu  -
{\sfrac 1 3}{\delta{\cal L}_\mu  \over \delta \Theta}\right) = \varepsilon$.
Hence, now the only possible additional requirement is the condition
(\ref{GACT}), which leads to the longitudinal torsion vector with respect
to the four-velocity $u_\alpha$
and to the local energy-momentum conservation law which is compatible with
the use of the {\em usual volume element} in its integral form.

As we see, all types of fluid's dynamics under consideration permit
a longitudinal torsion vector. Under this additional condition in all cases
the torsion force density vanish and we will reach the autoparallel equation
of motion for the fluid (\ref{AFluEM}) with the restriction
$\epsilon^{\alpha\mu\nu\lambda} u_\mu \partial_\nu u_\lambda \equiv 0$
on the four-velocity field $u_\alpha$. Hence, the autoparallel motion
for the fluid may be reached with the help of these additional requirements.

In the only case I.b when torsion vector may be transversal,
the torsion force density will have the form
${\cal F}_\alpha = (\varepsilon +p)\nabla_\alpha\Theta$.

The results of the present Section show that the additional conditions like
relations (\ref{AACT}), or (\ref{GACT}) which are needed in TEATG to ensure
the local conservation of the  energy-momentum of matter only, are possible from
algebraic point of view. Their physical consequences and their compatibility with
the full set of dynamical equations need further investigation.

\section{A Possible Interpretation of the Torsion\\ Potential $\Theta$ in TEATG}

The variant of fluid's dynamics described in Section 4.3 deserves a special
attention because it admits a new curious interpretation of TEATG. In it we
have to deal with two volume elements: the usual one $\dV_0 = d^4x \sqrt{|g|}$,
and the modified one $\dV=d^4x \sqrt{|g|}e^{-3\Theta}$.
The usual volume element is needed for calculations of integrals when we study
the conservation of the fluid's matter and energy-momentum.
The modified volume element is used {\em only}
for calculations of the corresponding {\em action integrals}
for geometric fields, for matter fields, and for particles
(See formulae (\ref{Atot}),(\ref{SFL}),(\ref{FLagrangian}),(\ref{New_spA})\/).
According to the articles \cite{Saa1} -- \cite{Saa5}
we have the same result for the other matter fields:
for gauge fields and for spinor fields.
The kinetic part of their lagrangians do not depend on the torsion potential
$\Theta$ (according to the MCP in the lagrangian of spinor fields one uses
the full affine connection in the space
${\cal M}^{(1,3)}\{g_{\alpha\beta}(x), \Gamma^\alpha_{\beta\gamma}(x)\}$,
but because of the specific structure of the kinetic part of this lagrangian
only the traceless part of the contorsion tensor
${\tilde K}^\alpha{}_{\beta\gamma}$ enters in it).
Just the same happens in action of the spinless fluid and
of the spinless particles if we choose the third variant of dynamics for them.
Then the formulae (\ref{LD}) show that in such a variant of TEATG
the torsion potential $\Theta$ will enter into the action of all matter fields
and particles under consideration uniformly --
via the multiplier $e^{-3\Theta}$ in the lagrangian densities
$\Lambda_M= e^{-3\Theta}\sqrt{|g|}{\cal L}_M$.
In the same manner it enters into the action of the geometric fields for which
$\Lambda_G= e^{-3\Theta}\sqrt{|g|}{\cal L}_G$,
but here exists an additional dependence on the potential $\Theta$,
because it appears in the lagrangian ${\cal L}_G$, too (See formula (\ref{LG})).

This situation calls for a new interpretation of the torsion potential $\Theta$
as a quantity which describes the space-time variations of the Plank "constant''
according to the law
\ben
\hbar (x) = \hbar_\infty  e^{3\Theta(x)},
\la{Plank}
\een
$\hbar_\infty$ being the Plank constant in vacuum far from
matter\footnote{In space-time with dimension D we will have
$\hbar (x) = \hbar_\infty  e^{(D-1)\Theta(x)}$.}.

Indeed, according to the first principles described
in \cite{Dirac}, \cite{Feynman},
we actually need lagrangians and action integrals to write down
the quantum transition amplitude in a form of Feynman path integral on the
histories of all fields and particles.
In the variant of TEATG under consideration it has the form:
\ben
\int{\cal D}\left(
\vbox to 12pt{} g_{\alpha\beta}(x),\Theta(x),\phi(x),x(t),...\right)
\exp\left({\frac 1 {\hbar_\infty}}\int d^4x (\Lambda_G + \Lambda_M)\right)=
\nonumber \\
\int{\cal D}\left(
\vbox to 12pt{}g_{\alpha\beta}(x),\Theta(x),\phi(x),x(t),...\right)
\exp\left({\frac 1 {\hbar_\infty}}\int d^4x e^{-3\Theta(x)} (L_G + L_M)\right).
\la{QA}
\een

Now it is obvious that in TEATG the very Plank constant $\hbar$ may be
included in the factor $e^{(D-1)\Theta(x)}$, but more important is the
observation that we must do this, because the presence of this {\em uniform}
factor in the formula (\ref{QA}) means that we actually introduce a local
Plank "constant'' at each point of the space-time.
Indeed, if the geometric field $\Theta(x)$ changes slowly in a cosmic scales,
then in the framework of the small domain of the laboratory we will see
an effective "constant":
$\hbar (x)\approx\hbar_\infty e^{3\Theta(x_{laboratory})}= const=\hbar$.

It can be easily seen that the Saa's model for geometric fields
$g_{\alpha\beta}$ and $\Theta(x)$ in vacuum is equivalent
to the Brans-Dicke theory \cite{BD}, \cite{Brans} in vacuum with parameter
$\omega= -{\sfrac 4 3}= -{\sfrac D {D-1}}$.
The corresponding Brans-Dicke scalar field $\Phi = e^{-(D-1)\Theta(x)}$
in vacuum replaces the $\Theta$ field in Saa's model.
It is well known that the solutions for the scalar field in Brans-Dicke theory
outside the matter go fast to a constant \cite{BD}, \cite{Brans}.
Hence, the same property will have the $\Theta$ field in Saa's model and
the value of this field far from matter is some constant $\Theta_\infty$
which may be incorporated in a natural way into the value of Plank constant.
If we do this, we may accept the value  $\Theta_\infty\equiv 0$ as
an universal asymptotic value of the $\Theta$ field outside the matter,
and the standard experimental value of the Plank constant approximately
as an asymptotic value $\hbar_\infty$ of the new field $\hbar(x)$.

This change of the physical interpretation of the theory is quite
serious and needs detailed consideration.
In the present article we shall give only some preliminary remarks.

1. The new interpretation make us free of the unpleasant necessity to deal
with two volume elements in the same theory and brings us back to
the "normal" volume element $\dV_0$. Then the true meaning of the Saa's
model will be in the highly nontrivial dynamics of the Plank field
$\hbar (x)$ described by the lagrangian ${\cal L}_G$ in
formula (\ref{LG}) and in its uniform interaction with matter fields.

Here an important remark concerning the mass terms in the lagrangians
and field equations of the massive fields is needed.
We shell illustrate it on the massive scalar field described in the Section 3.
The letter $m_\phi$ in the mass term in the lagrangian (\ref{SFL})
and in the corresponding field equation (\ref{SFE})
actually denotes the inverse Compton length $\lambdab^{-1}$
which in physical units is $\lambdab^{-1} = {{m c} \over \hbar}$,
$m$ being the
mass of the classical particle described by quantum wave equation (\ref{SFE}).
In the standard theory with Plank constant $\hbar$ we can chose the units
$\hbar = c = 1$ and this gives the possibility to denote the inverse
Compton length $\lambdab^{-1}$ as a ''mass" $m_\phi$.
In a theory with Plank field $\hbar(x)\neq const$ this will be not correct
and the best way to proceed is to turn back to the original physical notations.
This means to put in all formulae in Section 3
\ben
m_\phi = \lambdab^{-1}= {{m c} \over \hbar}=
{{m c} \over {\hbar_\infty}}\,e^{-3\Theta} =
\lambdab_\infty^{-1} \,e^{-3\Theta}
\la{m_phi_vh}
\een
and to correct the formulae in which the term
${\delta{\cal L}_M \over \delta \Theta}$,
or the derivatives of $m_\phi$ may enter.

Hence, in the model with variable Plank "constant" the mass terms of the matter
fields introduce an additional {\em universal} interaction of these fields with
the Plank field $\hbar(x)$ described by the equation (\ref{Plank}).
For all massive fields this interaction is defined via the formula
for the inverse Compton wave length as shown in the equation (\ref{m_phi_vh}).
For example in the lagrangian of the massive scalar field we will have
a term $\lambdab_\infty^{-2} \phi^2 \,e^{-6\Theta}$,
in the lagrangian of the Dirac spinor field $\psi$ --
a term $\lambdab_\infty^{-1} \bar\psi\psi \,e^{-3\Theta}$, etc.
This is an extension of the MCP for the case of variable Plank field
connected with the torsion of the specific affine connection we consider.

2. The essential variation of the field $\Theta$ (hence of the Plank field
$\hbar (x)$) may take place in scales of order of Schwarzschild radius $R_S$
of a given body. This means that we can expect some deviations of the laws of
standard quantum mechanics with constant parameter $\hbar$ at such small
scales and that it will be hard to see them at the scales of usual laboratory.
Such deviations may be essential only for the physics in small domains
around the center of the stars ($R_S \approx$ several kilometers),
around the center of galaxies ($R_S \approx 10^{11}$ kilometers),
(if the the matter's distribution does not smooth the variations of the
Plank field $\hbar (x)$), or in the cosmological models.
These are just the domains in which we are looking for a new physics
being pressed by experimental evidences.

3. If the speculation suggested in this Section of present article
really takes place, we must reconsider the existing attempts to quantize
gravity taking into account the physical meaning
and the role of the field $\Theta$.

4. The original Saa's idea was to interpret the field $\Theta$ as a
dilaton field, which at first appeared as a scalar partner of the tensor
field $g_{\alpha\beta}$ in the low energy limit of string theory.
The dilaton field causes difficulties in these theories up to now
(See \cite{Hammond1} -- \cite{Hammond3} and references therein).
The present interpretation may be useful in dilaton theory, too,
because it makes us free of the necessity
to consider this field as a matter field.
Moreover, in present theory the field $\Theta$ is incorporated
into the very geometry of the space-time in a definite way
and will not violate the equivalence principle
associated with the full affine connection \cite{F1}
in contrast to the dilaton field in usual interpretation of
the low energy limit of string theory.

5. If together with dilaton field $\Theta$ we consider the general case
of nonzero anty-symmetric contorsion $\tilde K_{\alpha\beta\gamma}$
in presence of spinor fields, we may share from string theory some more
information:   the field $\tilde K_{\alpha\beta\gamma}$ must be potential
and its potential is an anty-symmetric tensor $\Psi_{\mu\nu}$
\cite{Hammond1} -- \cite{Hammond3}. Then we can joint the present
theory of the field $\Theta$ and the interesting Hammond's theory of the
field $\Psi_{\mu\nu}$ described in \cite{Hammond1} -- \cite{Hammond3} and
in the series of additional papers by the same author.
This way we will reach some new theory of gravity and matter
with propagating torsion in Einstein-Cartan spaces
which may be able to overcome the old difficulties in ECTG
and obviously will be reach in new physical effects.

All these possibilities, as well as the consistency of such theory and its
relations with the physical reality at present are open problems under further
study. The corresponding results will be presented somewhere else.

\section{Trajectories of Classical Particles in TEATG
as a Semiclassical Limit of Wave Equation}

Finally in the present article we will explain the solution of
the consistency problem described at the end of the Introduction.
This solution depends on the model of the particle dynamics we accept
in TEATG.
The very derivation of the semiclassical limit of the autoparallel
equation (\ref{sfEA}) for the free wave motion in Einstein-Cartan space
${\cal M}^{(1,3)}\{g_{\alpha\beta}(x), \Gamma^\gamma_{\alpha\beta}(x)\}$
is universal. Independently of the particle dynamics it yields the eikonal
equation
\ben
g^{\alpha\beta} \partial_\alpha \varphi \partial_\beta \varphi =
\left({{mc}\over{\hbar}}\right) ^2
\la{EIC_}
\een
with an inverse Compton length
$\lambdab_\phi^{-1} = {{mc}\over{\hbar}} = m_\phi$ --
the "mass" of the scalar field $\phi$, $m$ being the mass of the classical
particle which correspond to this wave field according to quantum mechanics.

First we shall consider the two cases:
1) the strict Saa's model (See Section 4.1)
and 2) the model with the variable Plank field (See Sections 4.3, 6).
In them the particle dynamics is derived from
{\em the standard} variational principle.
Hence, the  Hamilton formalism and especially,
the Hamilton-Jacobi theory are well known.

1. In the strict Saa's model the classical equation of motion of the free
test particle (\ref{spEG}) derived from the lagrangian
$L_0 = -mc \sqrt{g_{\mu \nu} \dot x^\mu \dot x^\nu}$
is of geodesic type.
The canonical four-momenta which correspond to this lagrangian are
$p_{0\alpha} = -mc { {g_{\alpha\beta}\dot x^\beta} \over
{\sqrt{g_{\mu \nu} \dot x^\mu \dot x^\nu}}}$.
They are subject to the constraint
$g^{\alpha\beta} p_\alpha p_\beta = (mc)^2$.
Introducing the classical action function $S(x)$ related with the
four-momenta according to the formula $p_\alpha = \partial_\alpha S$
we obtain the Hamilton-Jacobi equation
\ben
g^{\alpha\beta} \partial_\alpha S \partial_\beta S = (m c)^2
\la{H-J0}
\een
which coincides with the eikonal equation (\ref{EIC_}) if $S=\hbar\varphi$.
As we see, the semiclassical limit of the wave mechanics of autoparallel type
in this case leads to the geodesic motion of the classical trajectories
in Einstein-Cartan space. Hence, this semiclassical limit is consistent
with the model under consideration.

2. In the case of the variable Plank field the particle trajectories are
derived from the lagrangian $L = e^{-3\Theta}L_0$,
See the equation (\ref{New_spA}).
The canonical four-momenta $p_\alpha= e^{-3\Theta}\,p_{0\alpha}$
which correspond to this lagrangian are subject to the constraint
$g^{\alpha\beta} p_\alpha p_\beta = (mc\,e^{-3\Theta})^2$.
Introducing the classical action function $S(x)$ related with the
four-momenta according to the standard formula $p_\alpha = \partial_\alpha S$
we obtain the Hamilton-Jacobi equation
\ben
g^{\alpha\beta} \partial_\alpha S \partial_\beta S = (m c\,e^{-3\Theta})^2
\la{H-J}
\een
which coincides with the eikonal equation (\ref{EIC_})
if $S=\hbar_\infty\varphi$
because the relation (\ref{Plank}) take place in this model.
Hence, the semiclassical limit of {\em the corresponding} autoparallel
wave equation (\ref{sfEA}) with mass $m_\phi$ given by the formula
(\ref{m_phi_vh}) is consistent with the classical mechanics in the model
with variable Plank field $\hbar(x)$.

We can invert these arguments and {\em derive} the formula (\ref{m_phi_vh})
from the requirement for consistency of the semiclassical limit of the wave
equation and the classical mechanics of particles in this model.
This way we see that the universal interaction between the field
$\Theta$ and the massive fields according to the equation (\ref{m_phi_vh})
is a consequence of this consistency requirement.

3. In the model described in the Section 4.2 an additional problems appear.
For the non-standard variational principle based on the commutation relation
(\ref{ComRel_T}) a consistent Hamilton formulation is not known.
Therefore at first glance it is impossible to construct
the Hamilton-Jacobi equation for some classical action
function and to relate it with the classical particle trajectories as usual.
This seems to make impossible the comparison between the semiclassical
limit (\ref{EIC_}) of the autoparallel equation of wave motion (\ref{sfEA})
and the corresponding autoparallel equation of motion (\ref{spEA}).

But for the special affine connection we use, the Saa's condition
(\ref{CC}) yields the gradient condition (\ref{GrC}) and brings us to the
situation described very recently in \cite{Kl7}: we may derive
{\em the same autoparallel equation} (\ref{spEA}) for the classical
trajectories using {the standard variational principle} with commutation
relation (\ref{ComRel})
for the auxiliary lagrangian $\tilde L = e^{-\Theta}\,L_0$.
Then the canonical momenta for the lagrangian
$\tilde L$ :\,\, $\tilde p_\alpha= e^{-\Theta}\,p_{0\alpha}$
are subject to the constraint
$g^{\alpha\beta} p_\alpha p_\beta = (mc\,e^{-\Theta})^2$ and
introducing the classical action function $S(x)$ related with them
according to the standard formula $\tilde p_\alpha = \partial_\alpha S$
we obtain the Hamilton-Jacobi equation
\ben
g^{\alpha\beta} \partial_\alpha S \partial_\beta S = (m c\,e^{-\Theta})^2.
\la{H-JA}
\een
It is important to stress that being introduced in the framework of the
{\em standard} Hamilton-Jacobi approach for
the  auxiliary lagrangian $\tilde L$
the action function $S(x)$ is related in a standard {\em geometric} way with
the trajectories of the {\em autoparallel} equation (\ref{spEA}) derived
from  the lagrangian $\tilde L$ via the usual variational principle.
Then this function  $S(x)$ will be connected with
the wave mechanics of the classical particles in usual way
because this depends on the geometry and the geometric relations
do not depend on the analytic way they are derived.
In other words the use of the  auxiliary lagrangian $\tilde L$ instead of
the original one $L_0$ is not essential for the geometric relations between
the function $S(x)$ and the trajectories of the autoparallel equation
(\ref{spEA}) in configuration space. Hence, in the present model we can accept
the equation (\ref{H-JA}) as a Hamilton-Jacobi equation for autoparallels.
Now, the experience we gained in the previous consideration shows that the
semiclassical limit (\ref{EIC_}) of the autoparallel wave equation (\ref{sfEA})
is consistent with the autoparallel equation (\ref{spEA}) if we put
$S=\hbar\varphi$ and impose the additional condition:
\ben
m_\phi = {{m c} \over \hbar}\,e^{-\Theta} = \lambdab^{-1}\,e^{-\Theta}
\la{m_phi_AA}
\een
The last condition is analogous to the condition (\ref{m_phi_vh})
in the model with variable Plank field and shows that
the consistency of the semiclassical limit of the autoparallel wave equation
(\ref{sfEA}) and the corresponding autoparallel equation of motion (\ref{spEA})
dictates a new universal interaction of the massive fields with the field
$\Theta$.
For example in the lagrangian of the massive scalar field we will have
a term $\lambdab^{-2} \phi^2 \,e^{-2\Theta}$,
in the lagrangian of the Dirac spinor field $\psi$ --
a term $\lambdab^{-1} \bar\psi\psi \,e^{-\Theta}$, etc.
This is the corresponding extension of the MCP for the model of
TEATG we consider here.

As we see, a proper interpretation of the TEATG permits us to
derive the autoparallel equation (\ref{spEA}) as a semiclassical limit
of the wave equation of autoparallel type (\ref{sfEA}).

We shall not describe here in details the solution of the consistency problem
for the semiclassical limit of the wave equation and classical motion
in the fourth variant of the particle dynamics in TEATG considered
in the present article.
It may be reached combining the ideas which gave the solution in
the previous two cases.

The considerations in this Section show that the semiclassical limit
of the wave equation (\ref{sfEA}) does not lead to
a preferable model of particle dynamics in the TEATG.
Instead, the requirement of the consistency of this limit with the
classical motion yields a definite universal interaction of the massive
fields with the torsion potential $\Theta$ which is different for each of
the variants of the particle dynamics\footnote{For the strict Saa's model
this requirement actually leads to an absent of such interaction.}.
Thus in the special type of Einstein-Cartan spaces we study,
this consistency requirement may be considered as {\em a new principle}
for construction of the theory and enables one to extend the MCP.

How to generalize the above semiclassical considerations, as far as the whole
Saa's model for affine theories of more general type then TEATG
is still an open problem which will be considered somewhere else.

\section{Appendix}

Here we give the basic properties of the special type of affine geometry
in the TEATG in presence only of spin zero matter.
In this case the condition ${\tilde K}_{\alpha\beta\gamma}\equiv 0$
implies $K_{\alpha\beta\gamma} = - 2 g_{\alpha [\beta} S_{\gamma]}$,
$S_{\alpha\beta\gamma}=K_{[\alpha\beta]\gamma}=S_{[\alpha} g_{\beta ] \gamma}$,
i.e. we reach a geometry with semi-symmetric affine connection described in
\cite{Schouten}, but  with the additional restriction (\ref{GrC}).
Hence, the torsion reduces to the torsion
vector, which in its turn is potential:
\ben
{S_{\alpha\beta}}^\gamma = S_{[\alpha}\delta_{\beta]}^\gamma =
\partial_{[\alpha}\Theta\delta_{\beta]}^\gamma .
\la{AT}
\een
For Cartan curvature tensor, Ricci tensor and scalar curvature
in D-dimensional space
${\cal M}^{(D)}\{g_{\alpha\beta}(x), \Gamma^\alpha_{\beta\gamma}(x)\}$
with such special type of geometry we have:
\ben
R_{\alpha\beta\mu\nu}&=&\stackrel{\{\}}{R}_{\alpha\beta\mu\nu} +
4g_{[\alpha[\nu}\left(\nabla_{\beta]} S_{\mu]} -S_{\beta]} S_{\mu]} +
{\sfrac 1 2}g_{\beta]\mu]} S_\sigma S^\sigma\right),
\nonumber\\
R_{\alpha\beta}&= &\stackrel{\{\}}{R}_{\alpha\beta} +
(D-2)\nabla_\alpha S_\beta +
g_{\alpha\beta}\nabla_\sigma S^\sigma +(D-1)g_{\alpha\beta}S_\sigma S^\sigma,
\nonumber\\
R&=& \stackrel{\{\}}{R} + 2(D-1)\nabla_\sigma S^\sigma +
D(D-1) S_\sigma S^\sigma.
\la{ACurvatures}
\een
Because of the zero nonmetricity condition
$\nabla_\alpha g_{\beta \gamma}\equiv 0$ we have the properties:
\ben
R_{\alpha\beta(\mu\nu)}= 0, \,\,\,\, {R_{\alpha\beta\sigma}}^\sigma = 0.
\la{AMC}
\een
The second of the equations (\ref{ACurvatures}) gives
$R_{[\alpha\beta]}=(D-2)\nabla_{[\alpha}S_{\beta]}$.
Then the relation (\ref{AT}) leads to symmetric Ricci tensor:
\ben
R_{[\alpha\beta]} = 0.
\la{ARicciSim}
\een
As a consequence the Einstein tensor turns to be symmetric:
\ben
G_{\alpha\beta}= R_{\alpha\beta} -{\sfrac 1 2} g_{\alpha\beta}R =
G_{\beta\alpha}
\la{AG}
\een
with trace $G=-{\sfrac {D-2} 2}R$ as in Riemannian space.
In addition we have the inverse relation
$R_{\alpha\beta}= G_{\alpha\beta} -{\sfrac 1 {D-2}} g_{\alpha\beta}G$
\footnote{Note that only in four dimensional space we have got a more
simple and symmetric relations $G=-R$, and
$R_{\alpha\beta}= G_{\alpha\beta} -{\sfrac 1 2} g_{\alpha\beta}G$.}.
One may represent the Einstein tensor and its trace
in another convenient form:
\ben
G_{\alpha\beta}&=& \stackrel{\{\}}{G}_{\alpha\beta}
+ (D-2)\left(\nabla_\alpha S_\beta-g_{\alpha\beta}\nabla_\sigma S^\sigma\right)
- {\sfrac {(D-1)(D-2)} 2} g_{\alpha\beta}S_\sigma S^\sigma
\nonumber\\
G &=& \stackrel{\{\}}{G}
- (D-1)(D-2) \nabla_\sigma S^\sigma
- {\sfrac {D(D-1)(D-2)} 2} S_\sigma S^\sigma.
\la{AG1}
\een

Taking into account the relation (\ref{AT}) and the general property of the
curvature tensor in spaces with semi-symmetric connection:
${R_{[\alpha\beta\mu]}}^\nu=2 \delta^\nu_{[\alpha}\nabla_\beta S_{\mu]}$ we reach
\ben
{R_{[\alpha\beta\mu]}}^\nu=0.
\la{ARId}
\een
At the end the Bianchi identity
\ben
\nabla_{[\gamma} {R_{\alpha\beta]\mu}}^\nu =
-2 S_{[\gamma} {R_{\alpha\beta]\mu}}^\nu
\la{Bianchi}
\een
in spaces with semi-symmetric affine connection
after some algebra leads to the important identity
\ben
\nabla_\sigma G^\sigma_\alpha + 2 G^\sigma_\alpha S_\sigma = 0.
\la{AGId}
\een
In the absence of torsion this is the well known identity of general
relativity which leads to the conservation of the energy-momentum tensor
of matter.
The role of the generalized identity (\ref{AGId}) in TEATG is not investigated
up to now. We study this problem in presence of spinless matter.
In this case the identity (\ref{AGId}) may be represented in a form
$\stackrel{\{\}}{\nabla}_\sigma G^\sigma_\alpha = 2 R^\sigma_\alpha S_\sigma$,
too.

The above described formulae show how much in general may differ
the space-time geometry of TEATG from the Riemannian geometry if
only spinless matter is present. For a more precise description
of this geometry we have to take into account the dynamical
equations of this theory.

\bigskip
\bigskip
\bigskip
\bigskip

\noindent{\Large\bf Acknowledgments}
\bigskip

\noindent
 The author is grateful to S. Yazadjiev for many useful discussions
 during the preparation of this article,
 and especially to A. Saa, R. T. Hammond, and J. G. Pereira who ware kind
 to sent to the author prints of their articles on the subject of the
 present paper.

 The author would like, too to express his thanks to the unknown referee for
 the suggestion to explain in the present article how the semiclassical
 approach yields particle trajectories from wave equation in TEATG
 and for the useful remarks concerning the literature references.

 The work has been partially supported by
 the Sofia University Foundation for Scientific Researches, Contract~No.~245/97,
 and by
 the Bulgarian National Foundation for Scientific Researches, Contract~F610/97.

\end{document}